\documentclass[aps,reprint,superscriptaddress,
nofootinbib,showpacs,amsmath,amssymb]{revtex4-1}

\usepackage{latexsym,graphicx,slashed,bm,dcolumn}

\newcommand{\be}[1]{\begin{equation}\label{#1}}
\newcommand{\beq}{\begin{equation}}
\newcommand{\eeq}{\end{equation}}
\newcommand{\beqn}[1]{\begin{eqnarray}\label{#1}}
\newcommand{\eeqn}{\end{eqnarray}}

\newcommand{\dub}[2]{\left(\begin{array}{c}{#1}\\{#2}
\end{array}\right)}

\newcommand{\mat}[4]{\left(\begin{array}{cc}{#1}&{#2}\\{#3}&{#4}
\end{array}\right)}

\renewcommand{\to}{\rightarrow}

\def\ee{\end{equation}}

\def\lsim{\raise0.3ex\hbox{$\;<$\kern-0.75em\raise-1.1ex
\hbox{$\sim\;$}}}
\def\gsim{\raise0.3ex\hbox{$\;>$\kern-0.75em\raise-1.1ex
\hbox{$\sim\;$}}}
\def\cal{\mathcal}

\def\cM{{\cal M}}
\def\cN{{\cal N}}
\def\cO{{\cal O}}

\def\tphi{\bar{\phi}}

\def\tf{\bar{f}}

\def\tq{\bar{q}}
\def\tl{\bar{l}}

\def\te{\bar{e}}
\def\tu{\bar{u}}
\def\td{\bar{d}}
\def\tnu{\bar{\nu}}

\def\lpr{l^\prime}
\def\phpr{\phi^\prime}

\newcommand{\dm}{\delta m}

\newcommand{\vect}[1]{\mbox{\boldmath$#1$}}

\def\dm{\epsilon} 

\def\rB{{\rm B}}
\def\rL{{\rm L}}

\def\bsig{\mbox{\boldmath $\sigma$} }

\begin{document}

\title{Anti-dark matter: a hidden face of mirror world}

\author{Zurab~Berezhiani}
\affiliation{Dipartimento di Fisica e Chimica, Universit\`a di L'Aquila, 67100 Coppito, L'Aquila, Italy} 
\affiliation{INFN, Laboratori Nazionali del Gran Sasso, 67010 Assergi,  L'Aquila, Italy}

\date{\today}


\begin{abstract}
$\rB$ and $\rL$ violating interactions of ordinary particles with their twin partticles from hypothetical mirror world  
can co-generate baryon asymmetries in both worlds in comparable amounts,    
$\Omega'_{\rm B}/\Omega_{\rm B} \sim 5$ or so.
On the other hand, the same interactions  induce the oscillation phenomena between the neutral particles 
of two sectors which convert e.g. mirror neutrons into our antineutrons. 
These oscillations are environment dependent and can have fascinating physical consequences. 
\end{abstract}

\begin{flushright} 
{\it \large In memoriam Vadim Kuzmin and Lev Okun } 
\end{flushright} 

\maketitle

There may exist a hidden sector of particles 
which is an exact replica of the observable particle sector, so that  
all ordinary (O) particles: the electron $e$, proton $p$, neutron $n$, photon $\gamma$, 
neutrinos $\nu$ etc. have invisible twins: $e'$, $p'$, $n'$, $\gamma'$, $\nu'$ etc.\ 
which are sterile to our interactions $SU(3)\times SU(2)\times U(1)$ 
but have their own gauge interactions $SU(3)'\times SU(2)'\times U(1)'$ 
with the same couplings.  Such a parallel sector, coined as mirror (M) world 
was introduced long time ago against parity violation: 
for our particles being left-handed (LH),  {\it Parity} can be interpreted as a discrete 
 symmetry which exchanges them with their right-handed (RH) twins from M sector \cite{Mirror}. 

Mirror atoms, invisible in terms of ordinary photons but gravitationally coupled  to our matter, 
constitute a viable candidate for dissipative dark matter,  
with specific implications  for the cosmological evolution, formation and structure of galaxies and stars 
etc. \cite{BCV,BCV2,M-stars,Alice}.
Although mirror particle physics is identical to ours, in the early universe M world  
should have a smaller temperature, $T' < T$. 
In particular, the Big Bang nucleosynthesis (BBN) bounds 
require $T'/T < 0.5$ \cite{BCV}, whereas a stronger limit  $T'/T< 0.3$ emerges from the cosmological 
data on the large scale structure  and cosmic microwave background  \cite{BCV,BCV2}.  
This can be realised  if after inflation O and M sectors  are (re)heated in non-symmetric way and   
then both systems evolve adiabatically,   
with $T' /T$  remaining nearly invariant in all epochs until today \cite{BCV}. 

A straightforward way to establish existence of mirror matter is the 
experimental search for oscillation phenomena between O and M particles. 
Any neutral particle, elementary (as e.g. neutrinos) or composite (as the neutron or hydrogen atom)  
can have mixing with its mass degenerate mirror twin 
which can be induced by the lepton and baryon number violating 
interactions between two sectors \cite{ABS,BB-nn'}. 
Remarkably,   these interactions can play a crucial role also  
in the early universe,  suggesting a cogenesis mechanism  
which can induce  comparable baryon asymmetries in both O and M worlds \cite{BB-PRL,Alice}.  

In this paper we show that this picture suggest  the  sign of mirror baryon asymmetry 
from which follows that the mixing occurs between neutral mirror particles and  our antiparticles.  
Hence, mirror world can be manifested experimentally as a hidden anti-world.

One can consider a theory based on the product $G\times G'$ of two 
identical gauge factors (Standard Model  or some its extension),   
O  particles belonging to $G$ and M  particles to $G'$,  
with the Lagrangian ${\cal L}_{\rm tot} = {\cal L} + {\cal L}' + {\cal L}_{\rm mix} $,   
where ${\cal L}_{\rm mix}$ stands for possible interactions between the particles 
of two worlds.
 The identical forms of the Lagrangians ${\cal L}$ and ${\cal L}'$ can be understood as 
a result of discrete symmetry $G\leftrightarrow G'$
when all O particles (fermions, Higgses and gauge fields) exchange places with their 
M twins (`primed' fermions, Higgses and gauge fields).
As we shall see below, 
such a discrete symmetry can be imposed {\it with} or {\it without} chirality change between 
the O and M fermions, with drastically different consequences.  

In the Standard Model $G=SU(3)\times  SU(2)\times U(1)$     
the fermions are represented as the Weyl spinors,  the LH ones $f_L$ 
 transforming as doublets of electroweak  $SU(2)\times U(1)$
and the RH ones $f_R$  as singlets (we omit family indices):
\be{SM-L} 
f_L: ~q_L = \dub{u_L}{d_L}, ~ l_L = \dub{\nu_L}{e_L};  
~~
f_R: ~u_R, ~ d_R, ~ e_R  
\end{equation} 
whereas anti-fermion fields $\tf_{R,L} = C \gamma_0 f_{L,R}^\ast$  
have  the opposite chiralities and opposite gauge charges: 
   \be{SM-R} 
\bar f_R: ~\tq_R = \dub{\tu_R}{\td_R}, ~ 
\tl_R = \dub{\tnu_R}{\te_R}; 
~
\tf_L: ~\tu_L, ~ \td_L, ~ \te_L 
\end{equation}
In addition, we prescribe a global baryon charge $\rB =1/3$  to 
quarks $q_L,u_R,d_R$, and a lepton charge $\rL=1$ to the leptons $l_L,e_R$. 
Then antiquarks $\tq_R,\tu_L,\td_L$ have $\rB=-1/3$, 
and antileptons $\tl_R,\te_L$ have $\rL=-1$.  

M sector,  with the gauge symmetry $G^\prime =SU(3)^\prime\times SU(2)^\prime\times U(1)^\prime$, 
has the analogous field content  
\be{SM-Lpr} 
f^\prime_L\!: ~q^\prime_L = \dub{u'_L}{d'_L}\!, ~ 
l^\prime_L = \dub{\nu'_L}{e'_L}\!;  
~~ 
f^\prime_R\!: ~u'_R, ~d'_R, ~e'_R  
\end{equation}
\be{SM-Rpr} 
\tf^\prime_R\!: ~ \tq'_R = \dub{\tu'_R}{\td'_R}\!, ~ 
\tl^\prime_R = \dub{\tnu'_R}{\te'_R}\!; 
~~ 
\tf_L\!: ~\tu'_L,~\td'_L,~\te'_L 
\end{equation}
For definiteness, let us precribe mirror fermion numbers: 
$\rB'=1/3$ to quarks $q'_L,u'_R,d'_R$ and $\rL'=1$ to leptons $l'_L,e'_R$. 
Then antiquarks $\tq'_R,\tu'_L,\td'_L$ have $\rB'=-1/3$, 
and antileptons $\tl'_R,\te_L$ have $\rL'=-1$.  
 
$\rB$ and $\rL$, related to accidental global symmetries possessed by the Standard Model 
at the level of  renormalizable Lagrangian terms,    
can be explicitly violated by higher order operators involving a large mass scale $M$. 
Namely, D=5 operator $\cO_5 = \frac{A}{M}(l\phi)^2 + {\rm h.c.}$ ($\Delta \rL=2$), 
$A$ being the coupling constants matrix in flavor space and $\phi$ the Higgs doublet,   
gives small Majorana masses to the neutrinos $\nu$, $m_\nu \sim \langle \phi\rangle^2/M$.   
Analogous operator $\cO'_5 = \frac{A}{M}(l'\phi')^2 + {\rm h.c.}$ ($\Delta \rL'=2$) 
gives masses to M neutrinos $\nu'$.  
But  there can exist also mixed  operator $\cO^{\rm mix}_5 =  \frac{D}{M}(l \phi)(l' \phi') + {\rm h.c.}$  
($\Delta \rL, \Delta \rL'=1$)  that induces $\nu - \nu'$ (active$-$sterile) mixings \cite{ABS}. 
In this way, mirror neutrinos,  being light on the same grounds as ordinary neutrinos,  
are natural candidates for  sterile neutrinos. 

These operators  can be induced in seesaw manner by introducing heavy gauge singlet neutrinos 
$N$ which can couple both $l$ and $l'$  and thus play the role of messengers between O and M sectors.  
The Yukawa Lagrangian 
\be{Yuk-l}
y \phi l N + y' \phi' l' N + g M N^2 + {\rm h.c.} 
\ee 
where $y,y'$ and $g$ are the matrices of Yukawa constants, leads to operators 
$\cO_5$, $\cO'_5$ and $\cO_5^{\rm mix}$,
with $A=y g^{-1} y^T$, $A'=y' g^{-1} y^{\prime T}$ and $D = y g^{-1} y^{\prime T}$. 

Analogously,  
D=9 operator $\cO_9 \sim \frac{a}{\cM^5}(udd)^2+{\rm h.c.}$ ($\Delta \rB=2$) 
induces mixings between the neutral baryons and anti-baryons, 
e.g. neutron$-$antineutron  ($n-\bar n$) mixing \cite{nnbar},  
similar operator $\cO'_9 \sim (a'/\cM^5)(u'd'd')^2+{\rm h.c.}$ ($\Delta \rB'=2$) acts in M sector, while 
the mixed one  
\be{n-npr}
\cO^{\rm mix}_9 \sim 
\frac{b}{\cM^5} (udd)(u'd'd')  +{\rm h.c.} \quad (\Delta \rB,\Delta \rB'=1) 
\ee 
leads to mixings $n-\bar n'$ etc.  
Also these operators can be seesaw-induced  by the exchange of some heavy singlet fermions $\cN$   
via the following Lagrangian terms \cite{BB-nn'}
\be{S}
S u d     +  S' u' d'  + \bar S d \cN  + \bar S^{\prime} d' \cN   +  M \cN^2 +   {\rm h.c.}   
\ee 
involving a color-triplet scalar $S$ with mass $M_S$ 
and its mirror partner $S'$ (coupling constants are suppressed). 
Integrating out the heavy states,
operators like (\ref{n-npr}) are induced effectively with  $\cM^5 \sim M_S^4 M  $. 
Taking e.g. $M_S\sim 1$ TeV and  $M \sim 10^{13}$ GeV, one gets $\cM \sim 100$~TeV. 

The same $\rB$ or $\rL$ violating interactions between O and M particles 
can also generate baryon asymmetries in both sectors, via processes 
that transform ordinary leptons or quarks into mirror ones in the Early Universe.  
Let us discuss e.g. the co-leptogenesis scenario  \cite{BB-PRL}
via the scattering processes $l\phi \to \tl'\tphi' $ etc. due to interaction terms (\ref{Yuk-l}). 
(Alternatively, one could discuss co-baryogenesis scheme via processes $d \bar S \to \bar d' S'$
due to couplings (\ref{S}) \cite{BB-nn'}.) 

Let us assume, for simplicity, that after inflation only O world heats up,  
$T_R$ being the reheating temperature, while M sector 
is almost ``empty" at the beginning, $T'=0$ (imagine e.g. the chaotic inflation picture 
with two inflation scalars, where one is excited and after inflation it decays into O particles, 
while another (mirror) field has a small initial value). 
We assume also that the masses $N$ fermions are much larger than reheating temperature,  
$M\gg T_R$,  so that they  never appear in the thermal bath  
but  they mediate processes like $l\phi \to \tl'\tphi' $ which heats up also M sector.  
An imaginary astronomer living in the epoch $T\sim T_R$  
would observe an additional (to the Hubble expansion) cooling of O world  
due to the particle  leakage to parallel world, and  production of non-zero $B-L$ 
as far as the leptons $l$ and antileptons $\bar{l}$ leak with different rates.  
On the other hand, his mirror colleague would observe the entropy production in M world,
with the leptons $l'$ and antileptons $\bar{l}'$ emerging with different rates so that   
a non-zero $\rB'-\rL'$ is induced at the end.
In fact, all three conditions for baryogenesis \cite{Sakh,KRS} are fulfilled 
since these processes violate $\rB-\rL$  in both sectors, violate CP  
due to complex couplings $y$ and $y'$ in (\ref{Yuk-l}), and 
they are out of equilibrium (since after their freezing 
two sectors should end up with different temperatures, $T'/T< 0.3$ or so \cite{BCV,BCV2}).  

The evolution of $\rB-\rL$ and $\rB'-\rL'$ number densities 
are described by the equations \cite{BB-PRL,Alice}
\beqn{L-eq}
&& {{ d n_{\rm BL} } \over {dt}} + (3H + \Gamma) n_{\rm BL} 
= \frac34 \Delta\sigma \, n_{\rm eq}^2 ,  \nonumber \\
&& {{ d n'_{\rm BL} } \over {dt}} + (3H + \Gamma' ) n'_{\rm BL}  =
  {3 \over 4} \Delta\sigma' \, n_{\rm eq}^2   \; ,
\eeqn
where $n_{\rm eq}\simeq(1.2/\pi^2)T^3$, 
$ \Gamma \simeq {\rm Tr}(A^\dagger A)n_{\rm eq} /M^2$ and 
$\Gamma' \simeq {\rm Tr}(A^{\prime \dagger} A')(T'/T)^3 n_{\rm eq}/M^2$ 
are $\Delta\rL,\Delta\rL=2$ reaction rates in two sectors,  
while $\Delta\sigma \approx 0.2 JT^2/M^4$   
($J= {\rm Im\, Tr} [ g^{-1}(y^{\dagger}y)^\ast g^{-1}(y^{\prime\dagger}y^\prime) g^{-2} (y^\dagger y) ] $ being 
the CP-violating factor) and $\Delta\sigma'$ 
(obtained from $\Delta\sigma$ by exchange $y\leftrightarrow y'$) 
  take into account that the processes with $l$ and $\tl$ 
in the initial state have different (thermally averaged) cross-sections     
due to the interference between the tree-level and one-loop diagrams 
shown in Ref. \cite{BB-PRL}: 
\begin{eqnarray}\label{CP}
&&
\sigma (l\phi\to \tl'\tphi') -
\sigma(\tl\tphi \to \l'\phi') =
(- \Delta\sigma  - \Delta\sigma' ) /2
\, ,  \nonumber \\
&&
\sigma (l\phi\to \lpr\phpr) -
\sigma(\tl\tphi \to \tl'\tphi') =
( -\Delta\sigma + \Delta\sigma' )/2
\, ,   \nonumber  \\
&&
\sigma (l\phi\to \tl\tphi) -   
\sigma(\tl\tphi \to l\phi) = \Delta\sigma \, . 
\end{eqnarray}

Until now we left the Yukawa coupling constants $y$ and $y'$ in (\ref{Yuk-l}) as independent parameters. 
Let us impose discrete mirror symmetry ${\cal P}(G\leftrightarrow  G')$ 
under the exchange between the particle sets (\ref{SM-L})--(\ref{SM-R})  and (\ref{SM-Lpr})--(\ref{SM-Rpr}) as
\be{MLR}
{\cal P}: ~ f_{L(R)}  \leftrightarrow \tf'_{R(L)},  \quad \tf_{R(L)} \leftrightarrow f'_{L(R)}, 
\quad  \phi \leftrightarrow \tphi' , 
\end{equation}
complemented with adequate exchange of O and M gauge fields.  
 ${\cal P}$ can be viewed as a generalized {\it Parity} transformation which exchanges the LH and RH fermions   
while keeps invariant combined fermion charges $\bar \rL = \rL-\rL'$ and $\bar \rB = \rB-\rB'$,   
thus restoring the left-right symmetry between O and M Lagrangians  
right in the spirit of historical papers \cite{Mirror}. 
For the quark and lepton Yukawa constants of two sectors it implies that $y'_{u,d,e} = y^\ast_{u,d,e}$, 
i.e. the same CP-violating phases that our {\it particles} have in the LH basis (\ref{SM-L}),  
mirror sector has in the RH ({\it antiparticle}) basis (\ref{SM-Rpr}). 
Hence, should primordial baryogenesis in two sectors occured independently, 
e.g. via scenarios discussed in Ref. \cite{BCV}, 
one would expect their baryon asymmetries having the opposite signs. 

However, in our cogenesis mechanism right the opposite takes place.  
The  invariance under  
$l \to \bar l'$, $N \to \bar N$, $\phi \to \bar \phi$ implies $y'=y^\ast$  and $g^\ast = g$ 
in (\ref{Yuk-l}), then $\Delta\sigma' =\Delta\sigma$ in Eqs. (\ref{L-eq}),  
and so O and M baryon asymmetries must have the same sign:   
once $B > 0$,  then also $B' > 0$.  
Hence, the left-right Parity ${\cal P}$ between the Lagrangians of two worlds  is violated in the realisation 
of the universe: O and M worlds both appear to be left-handed.\footnote{
In alternative, by imposing discrete symmetry ${\cal C}(G\leftrightarrow G')$ 
without changing charities:  
$f_{L,R} \leftrightarrow f'_{L,R}$,  $\tf_{R,L} \leftrightarrow \tf'_{R,L}$, $\phi \to \phi'$,   
we get  $y'=y$ and so  $\Delta\sigma$ and $\Delta\sigma'$ are vanishing.   
This makes our cogenesis mechanism ineffective.  }

Solving Eqs. (\ref{L-eq}), one gets the cosmological fraction of dark baryons 
$\Omega_B' h^2 \simeq 10^3J M_{Pl} T_R^3/M^4$ \cite{BB-PRL,Alice}   
which has the right value   
e.g.  for $M\sim 10^{13}$ GeV, $T_R \sim 10^{11}$ GeV and $J \sim 10^{-3}$. 
In O world, since it  is hotter,  the baryon asymmetry can have some dumping.   
In fact, one has  $0.2 < \Omega_B/\Omega'_B < 1$, where 
the lower bound emerges by demanding $T'/T < 0.3$ for the final temperatures 
while the upper one corresponds to $T'/T \to 0$ \cite{Alice}. 

Besides operators like ${\cal O}_5^{\rm mix}$ and ${\cal O}_9^{\rm mix}$, the Lagrangian $\cal L_{\rm mix}$ 
can contain terms like kinetic mixing of photons $\frac{\varepsilon}{2} F^{\mu\nu} F'_{\mu\nu}$ \cite{Holdom} 
or some common gauge interactions between two sectors  \cite{PLB-su3}. 
In particular, the photon kinetic mixing induces the Coulomb potential 
$V(r) = \varepsilon \alpha ZZ'/r$ between O and M particles with respective electric charges $Z$ and $Z'$.  
Cosmological bounds allow  $\varepsilon \sim 5\cdot 10^{-9}$ or so \cite{Lepidi}, 
compatible also with the limits from the direct detection experiments of dark matter.  
Then,  a significant amount of mirror particles passing through the sun 
can be captured due to Rutherford-like scatterings, 
with total captured mass of about $10^{-12} M_\odot$, or even bigger   
since M gas density in the galactic disk can be larger than 1 GeV/cm$^3$. 
This can improve the agreement of the solar models with helioseismic data \cite{Panci}. 
 Most of the captured M matter will settle in the central part of the sun, while some fraction 
 of it can form an extended cloud populated by slowly moving M particles, 
with velocities up to 40 km/s at the Earth orbit.  
Then, taking their density as e.g. 100 GeV/cm$^3$ at the distances $\sim 1$ AU from the sun,  
which is compatible with the upper limits of the dark matter density in solar system \cite{Yulik},   
 also the Earth can capture substantial amount of M matter, 
$\sim 10^{-6} M_\oplus$, via the same Rutherford-like scatterings.  

Interestingly, the photon kinetic mixing  can give rise also to mirror magnetic fields on the Earth,  
via the mechanism suggested in Ref. \cite{BDT}. Provided that 
captured M matter is partially ionised (say $10^{-5}$ part of it),  
the drag force exerted by the Earth rotation on free M electrons   
induce circular electric currents 
sourcing the mirror magnetic fields as
$B' \sim \varepsilon^2 \times 10^{15}  $~G, which value can be further amplified  
by the dynamo mechanism. 

 Let us turn now to the oscillation phenomena between two sectors.    
Operator (\ref{n-npr})  induces the mass mixing 
$\dm n n' + {\rm h.c.}$ ($\bar \rB = \rB-\rB'$ is conserved) which 
gives rise to  the neutron oscillation into mirror antineutron, 
 experimentally  detectable as the neutron disappearance $n\to \bar n'$ 
 as well as regeneration  $n\to \tilde n' \to n$. 
 M neutrons, vice versa, should oscillate into our antineutrons, $n' \to \bar n$.   
 As it was pointed out in Ref. \cite{BB-nn'}, $n\to \bar n'$  oscillation time, 
 $\tau_{n\bar n'} = \dm^{-1} \sim (\cM/10~{\rm TeV})^5$~s,  
 can be much smaller  
 than the neutron decay time;  
 in fact, $\tau_{n\bar n'} \sim 1$ s or even less is allowed by the experimental and astrophysical bounds. 
  Two moments are important: in difference from $n-\bar n$ oscillation \cite{Phillips}, 
  $n\to \bar n'$  oscillation is ineffective for neutrons bound in nuclei and cannot destroy stable elements, 
 while   for free neutrons it is suppressed by the matter and magnetic field effects. 
In a generic medium $n-\bar n'$ oscillation is described by the effective Hamiltonian  
\be{H-osc}
H = 
\mat{ \mu \vect{B} \bsig + V_n - iW_n}{\dm} {\dm} { -\mu \vect{B}'  \bsig + V_{\bar n'}  - iW_{\bar n'} } 
\ee
where $\bsig=(\sigma_x,\sigma_y,\sigma_z)$ are the Pauli matrices, 
$\mu$ is the neutron magnetic moment, 
$\vect{B}$ and $\vect{B}'$ are magnetic fields  while   $V_n$, $V_{\bar n'}$ 
and  $W_n$, $W_{\bar n'}$ stand for the coherent scattering and absorption of $n$ and $\bar n'$ 
 in O and M media respectively. 
 In cosmic space $n\to \bar n'$ oscillation can have maximal probability: 
e.g. for  $\tau_{n\bar n'} \sim 1$ s,  it would suffice to have matter densities $< 10^{-8}$ g/cm$^3$ 
and magnetic fields $< 10^{-3}$ G in both media.    
In terrestrial experiments  M magnetism and M gas density cannot be suppressed  
but our magnetic fields can be tuned close to the resonance conditions 
for enhancing the oscillation probability. 
Several experiments studied the magnetic field dependence of the ultra cold neutron (UCN) loses 
\cite{Serebrov}. 
Interestingly, the measurements performed  at $B\simeq 0.2$ G 
 indicate to anomaly  at about $5\sigma$ away from the null hypothesis  \cite{Nesti}
which can be interpreted via $n\to \bar n'$ oscillation 
 in the presence of mirror field $B' \sim 0.1$ G 
with $\tau_{n\bar n'}\sim 2-10$ s. 
 
The presence of M matter can also have interesting implications 
for experimental measurements of the neutron lifetime. 
Oscillation $n\to \bar n'$ and annihilation of $\bar n'$  on mirror gas 
lead  to continuous UCN loses imitating the neutron decay,  
with the rate which can be estimated as $P_{n\bar n'} W_{\bar n'} \simeq 
(2~{\rm s}/\tau_{n\bar n'})^2 (\rho'/10^{-8}~ {\rm g\cdot cm}^{-3}) \times 10^{-6}~{\rm s}^{-1}$, 
where $P_{n\bar n'}$ is the average oscillation probability and 
$\rho'$ is the mirror gas density at the Earth surface. This can contribute  
as few seconds in the UCN storage time. For capturing this effect, 
the neutron lifetime should be measured  in magnetic fields of different strength. 

In neutron stars (NS) $n-\bar n'$ oscillation can be described by Hamiltonian (\ref{H-osc}) 
with $V$ taken as the neutron Fermi energy,  ${\cal E} \simeq (N/N_{\rm nuc})^{2/3} \times 60$~MeV, 
with $N/N_{\rm nuc}$ being the neutron number density in units of nuclear density.       
Then the  average oscillation probability   between the neutron 
collisions is $P_{n\bar n'} = (\dm/\cal E)^2$. 
Taking the collision frequency as 
$\sigma_{nn} v N \simeq 10^{24} \times (N/N_{\rm nuc})^{4/3}$  s$^{-1}$ 
(Pauli blocking is neglected),  the rate of $\bar n'$ production can be estimated as 
$\Gamma \simeq (1~{\rm s}/\tau_{n\bar n'})^2   \times 10^{-15}$ yr$^{-1}$. 
The produced  $\bar n'$ decay as $\bar n' \to \bar p' \bar e' \nu'_e$ and relativistic $\bar e'$  
escape from the star while $\bar p'$  also evaporate by the electric repulsion 
 since no more than $10^{21}$ charged particles can be contained.
 Hence, the mass of the NS should decrease by time as $M \sim \exp(-\Gamma t) M_0$. 
 For $\tau_{n\bar n'} \sim 1$ s, $\Gamma^{-1}$  is much larger than  the age of the universe. 
 But for $\tau_{n\bar n'} \sim 10^{-3}$ s one would have $\Gamma^{-1} \sim 10^{9}$ yr,  
 so the lightest and oldest NS  reaching the minimal mass limit today could 
  end up with  spectacular explosions giving rise to the phenomena like gamma ray bursts.   
 
Let us discuss the same process in mirror NS, with $n'$ 
transforming into our antineutron $\bar n$. 
The latter decays as $\bar n \to \bar p e^+ \nu_e$, and produced $e^+$  with energies $\sim 1$ MeV 
escape from the star.  Hence, mirror NS can be seen as cosmic engines producing positrons 
with the rate $(1~{\rm s}/\tau_{n\bar n'})^2  \times 10^{42}$ yr$^{-1}$, and about $10^5$  mirror NS could 
produce the positron amount  sufficient for explaining the 511 keV gamma excess 
from the galactic bulge.

In the BBN epoch $n\to \bar n'$ oscillation is suppressed by matter density \cite{BB-nn'}. 
But for relatively small $\tau_{n\bar n'}$, a moderate injection of antineutrons  
 due to $n' \to \bar n$ oscillation could help for settling the low deuterium and lithium problems. 
 This  question  deserves a special numerical study. 

 Free mirror neutrons can be produced by disintegration of M helium and heavier M nuclei 
 in galactic cosmic rays by mirror ultraviolet light or mirror gas.   
 These  $n'$ will oscillate into our antineutrons and then decay as $\bar n \to \bar p e^+ \nu_e$. 
The antiprotons produced in this way would have same spectral index as progenitor cosmic rays,  
as hinted by the AMS2 measurements. 
 Oscillations $n-\bar n'$ and $n' \to \bar n$ can have strong effects also for ultra high energy cosmic rays,  
 GZK cutoff and cosmogenic neutrinos \cite{nn'-cosmic}. Unfortunately, at these energies  cosmic protons and 
 antiprotons cannot be distinguished. 
 
  Oscillation of mirror hydrogen atoms into our anti-hydrogen via operator 
 $ \frac{1}{M^8}(uude)(u'u'd'e')$,  with reasonably small mass scale $M\sim 1$ TeV 
 and thus oscillation time $\sim 10^{20}$ s, can also have  interesting implications 
 for galactic gamma background.

Concluding, we have shown that mirror matter can be considered as anti-dark matter, 
or grey antimatter, since M particles with some probabilities can be converted  into our antiparticles. 
These phenomena, subject to the tuning of environmental conditions,  
can have fascinating phenomenological and astrophysical implications. 
  
However, they can also have far going consequences which by now were 
touched only in science fiction:  exchange of matter with parallel world could provide an alternative source 
of the energy,  as envisaged by Isaac Asimov in {\it ``The Gods Themselves"} in 1972.  
Imagine e.g. that mirror physicists built a reactor in their cosmic station, while in its vicinity 
we construct a big chamber where the magnetic fields  are tuned so that $n' \to \bar n$ 
transition occurs in resonance regime (we could communicate with M aliens via electromagnetic 
kinetic mixing). Then, if oscillation time is indeed short, 
$\tau_{n\bar n'} \sim 1$~s  or so \cite{Nesti}, 
a significant part of produced $n'$  could be converted into our antineutrons  
giving a lot of energy after their annihilation, about $10^3$ times more than produced by reactor.

\bigskip 


\noindent 
I feel privileged for having many illuminating conversations with Vadim  Kuzmin and Lev Okun 
on mirror world and mirror neutrons. 
Interesting discussions with Annabel Berejiani and Yuri Kamyshkov are also acknowledged.  
This work was supported in part by MIUR grant on "Astroparticle
Physics" PRIN 2012CPPYP7.


\end{document}